\documentclass[reprint, aps,superscriptaddress,amsmath,amssym,prl]{revtex4-2}

\usepackage{bm}
\usepackage{float}
\usepackage{graphicx}
\usepackage[usenames,dvipsnames]{color}
\usepackage[normalem]{ulem}
\usepackage[svgnames]{xcolor}
\usepackage{bm}% bold math
\usepackage{multirow}
\usepackage{titlesec}
\usepackage[utf8]{inputenc}
\usepackage[colorlinks,linkcolor=blue,citecolor=blue,urlcolor=blue]{hyperref}
\usepackage{booktabs}
\usepackage{array}
\usepackage{url}
\usepackage{tabularray}

\begin{document}

\title{Electron-phonon coupled hydrodynamics in semimetal TaAs$_2$}

\author{Shan Jiang}
\affiliation{Wuhan National High Magnetic Field Center and School of Physics, Huazhong University of Science and Technology,  Wuhan  430074, China}
\affiliation{Laboratoire de Physique et d'\'Etude de Mat\'{e}riaux (CNRS)\\ ESPCI Paris, PSL Research University, 75005 Paris, France }

\author{Wei Xie}
\affiliation{Wuhan National High Magnetic Field Center and School of Physics, Huazhong University of Science and Technology,  Wuhan  430074, China}

\author{Xiaokang Li}
\affiliation{Wuhan National High Magnetic Field Center and School of Physics, Huazhong University of Science and Technology,  Wuhan  430074, China}

\author{Kamran Behnia}
\email{kamran.behnia@espci.fr}
\affiliation{Laboratoire de Physique et d'\'Etude de Mat\'{e}riaux (CNRS)\\ ESPCI Paris, PSL Research University, 75005 Paris, France }

\author{Zengwei Zhu}
\email{zengwei.zhu@hust.edu.cn}
\affiliation{Wuhan National High Magnetic Field Center and School of Physics, Huazhong University of Science and Technology,  Wuhan  430074, China}

\date{\today}

\begin{abstract}

Hydrodynamic corrections to diffusive transport can arise when momentum-conserving collisions between quasiparticles become prominent, and they have been documented for both electrons and phonons. An emerging frontier topic is coupled electron-phonon ($e$-$ph$) hydrodynamics. Here, through electrical and thermal transport measurements on TaAs$_2$ crystals with different impurity levels, we document the emergence of an $e$-$ph$ bifluid in the temperature window of 5 to 15~K. Within this range, the lattice thermal conductivity exhibits a faster-than-$T^3$ temperature dependence, as a consequence of non-monotonic and purity-dependent phonon mean free paths, a signature of phonon Poiseuille flow. However, strong $e$-$ph$ coupling impedes the emergence of a ballistic regime. This is corroborated by the observation of quantum oscillations in the lattice thermal conductivity. Prominent phonon-mediated momentum exchange between electrons amplifies the violation of the Wiedemann-Franz law and yields a two-order-of-magnitude discrepancy between quantum and transport lifetimes, a signature of electron hydrodynamics in semimetals. Our results imply that in semimetals with optimized $e$-$ph$ coupling, thanks to matching between the cryogenic phonon wavelength and the Fermi wavelength, momentum and energy flow between the electron and phonon reservoirs as frequently as within each reservoir.

\end{abstract}

 \maketitle

\section{INTRODUCTION}

Over fifty years ago, Gurzhi predicted that when momentum-conserving scattering dominates among electrons in metals, or phonons in insulators, hydrodynamic behavior emerges, modifying charge or entropy transport according to the laws of viscous fluid mechanics~\cite{Gurzhi1968}. However, most collisions among quasiparticles in a solid are of the Umklapp type and dissipate momentum. Nevertheless, signatures of phonon hydrodynamics have been observed in several crystalline insulators~\cite{Beck1974, Martelli2018, Machida2018, Machida2020, kawabata2025}, and signatures of electron hydrodynamics have been reported in a limited number of ultra-clean metals~\cite{deJong1995, Bandurin2016, Crossno2016, krishna2017, guo2026hydrodynamicsviscouselectronfluid}.

A subject of recent attention are cases where there is a significant exchange of momentum between the electronic and phononic reservoirs. It has been proposed that phonon-mediated electron-electron ($e$-$e$) scattering may drive electron hydrodynamics in semimetals~\cite{Coulter2019, Vool2021, Osterhoudt2021, Wang2022}, as well as $T$-linear resistivity in high-temperature superconductors above their critical temperature~\cite{Davison2014, Lucas2017}. Additionally, the momentum and energy exchange between the electron and phonon baths has been proposed to give rise to a coupled electron-phonon ($e$-$ph$) bifluid~\cite{Levchenko2020}.

Semimetals have emerged as a highly promising platform for such studies. Their small Fermi radius weakens the probability of Umklapp $e$-$e$ scattering and, by matching the acoustic phonon wavelength at cryogenic temperatures, strengthens the $e$-$ph$ coupling. Transport studies on WP$_2$~\cite{Gooth2018}, Sb~\cite{Jaoui2021}, and WTe$_2$~\cite{Xie2024} have revealed a large downward deviation from the Wiedemann-Franz (WF) law, an effect attributed to electron hydrodynamics and further reinforced by $e$-$ph$ coupling~\cite{Jaoui2022}. Magnetic imaging of the electron fluid in WTe$_2$~\cite{Vool2021} has identified a Poiseuille profile of electron flow and linked it to $e$-$ph$ coupling. Calculations~\cite{Huang2021, Levchenko2020} have put $e$-$ph$ hydrodynamics on a sound theoretical basis. Further support for this picture has been provided by Raman scattering experiments in WP$_2$~\cite{Osterhoudt2021}. Nevertheless, transport evidence for $e$-$ph$ hydrodynamics has remained elusive.

\begin{figure*}[ht]
\centering
\includegraphics[width=15cm]{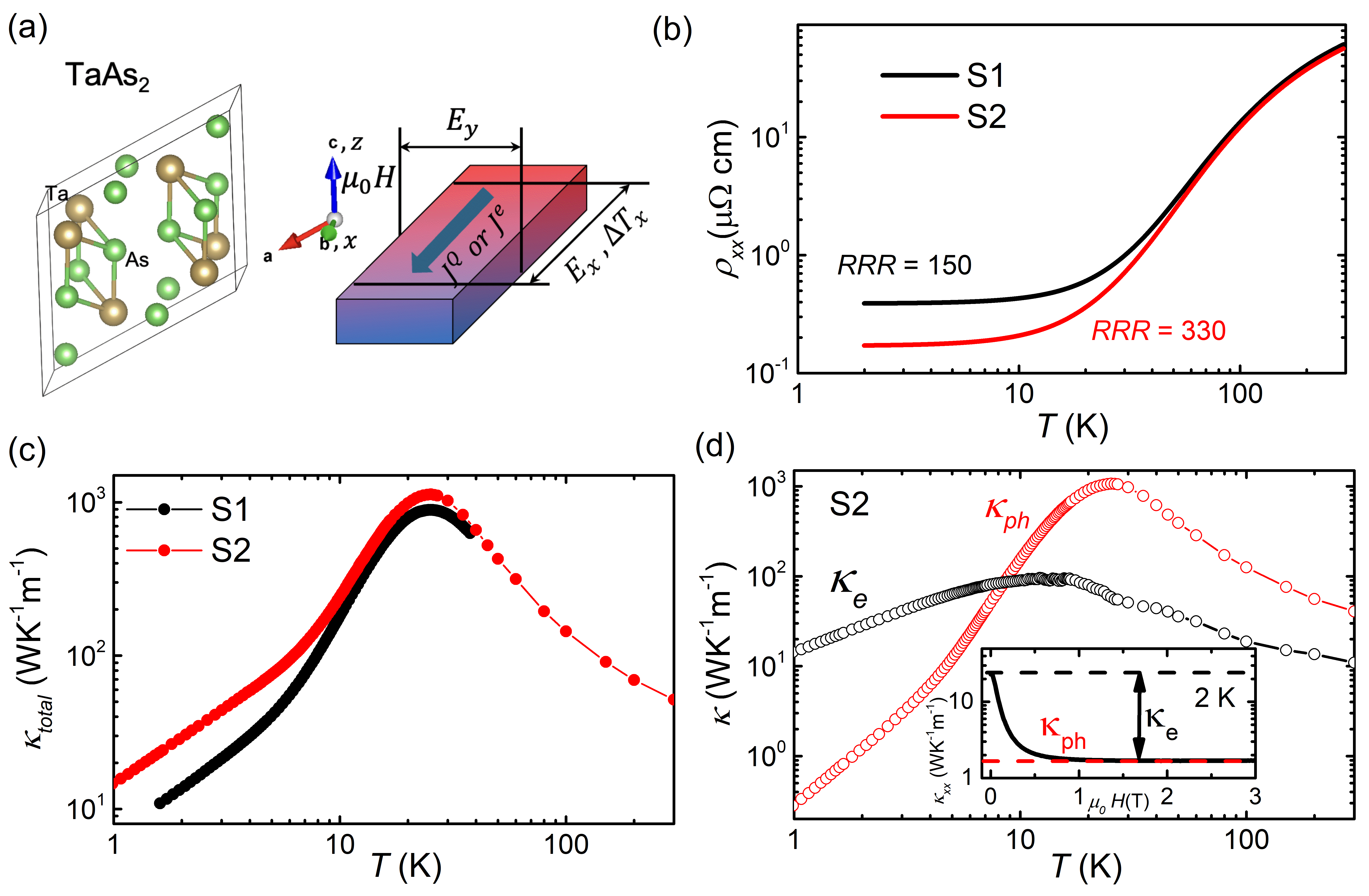}
\caption{\textbf{Electrical resistivity and thermal conductivity in TaAs$_2$.}
(a) Schematic of the transport measurement configuration. Heat and current are applied along the $b$-axis, with the magnetic field oriented parallel to the $c$-axis. Right panel: projection of the crystal structure onto the $ac$-plane.
(b) Temperature dependence of the electrical resistivity $\rho_{xx}$ for two TaAs$_2$ samples with different residual resistivity ratios ($RRR = \rho_{300\mathrm{K}}/\rho_{2\mathrm{K}}$). Inset: temperature exponent $\gamma$ extracted from $\rho = \rho_0 + A_{\gamma}T^{\gamma}$. At low temperatures, $\gamma \approx 2$, indicating Fermi liquid behavior; it increases to a maximum of $\sim 2.8$ owing to $e$-$ph$ scattering.
(c) Temperature-dependent total thermal conductivity for the two samples.
(d) Temperature dependence of the lattice ($\kappa_{\mathrm{ph}}$) and electronic ($\kappa_{\mathrm{e}}$) thermal conductivity, extracted under an applied magnetic field (see Supplementary Information\cite{SM}). Inset: field-dependent total thermal conductivity $\kappa_{xx}$. The magnetic field suppresses $\kappa_{\mathrm{e}}$, leaving the phononic contribution $\kappa_{\mathrm{ph}}$ dominant.}
\label{fig1}
\end{figure*}

In this study, by measuring electrical and thermal transport in two single crystals of TaAs$_2$~\cite{Liu2020, Wadge2022}, we found signatures of hydrodynamic transport accompanying the previously reported ones. The electron and hole densities in this semimetal are $n \approx p \approx 7 \times 10^{18}$~cm$^{-3}$ (see two-band fitting of the electrical transport in the Supplementary Information~\cite{SM}). This means that each mobile electron and hole is shared by several thousand atoms. The high mobility of carriers ($10^{5}$~cm$^{2}$~V$^{-1}$~s$^{-1}$) leads to a large magnetoresistance, allowing us to employ a magnetic field~\cite{White01041958, Jaoui2021, Gourgout2024} in order to separate the electronic and phononic contributions to thermal conductivity (see Supplementary Information~\cite{SM}).

We find that hydrodynamic features associated with both phonons and electrons emerge within a narrow temperature window between 5 and 15~K. The lattice thermal conductivity $\kappa_{\mathrm{ph}}$ exhibits a faster-than-$T^3$ temperature dependence, reflecting a non-monotonic, purity-dependent phonon mean free path, which is diagnosed as a signature of Poiseuille flow. However, the ballistic regime is never attained owing to strong $e$-$ph$ coupling, as evidenced by the observation of quantum oscillations in the lattice thermal conductivity. For electronic transport, the ratio of electrical to thermal conductivity falls well below the Wiedemann-Franz (WF) law expectation, originating from $e$-$e$ scattering. Moreover, the quantum lifetime is found to be two orders of magnitude shorter than the transport scattering time, implicating frequent momentum-conserving (MC) phonon-mediated $e$-$e$ scattering that gives rise to electron hydrodynamics, a scenario previously suggested for WTe$_2$~\cite{Vool2021}. All these observations point to $e$-ph hydrodynamics, where, thanks to the closeness of the Fermi and phonon wavelengths, momentum and energy flow as frequently as any flow inside each bath.

\section{RESULTS}

\noindent\textbf{Electrical and Thermal Transport Measurements.} 
TaAs$_2$ has a monoclinic structure with $\beta$ around $120^\circ$~\cite{Luo2016}. Heat or electric current is applied along the $b$-axis with the magnetic field parallel to the $c$-axis, as shown in Figure~\ref{fig1}(a). Figures~\ref{fig1}(b) and (c) display the resistivity and total thermal conductivity of two TaAs$_2$ samples. The $RRR$ ratios ($\rho_{300\mathrm{K}}/\rho_{2\mathrm{K}}$) of samples S1 and S2 are 150 and 330, respectively, reflecting different impurity levels, and the total thermal conductivity also exhibits a corresponding impurity dependence. As shown in the inset of Figure~\ref{fig1}(d), a magnetic field can effectively suppress the electronic thermal conductivity, leaving the phononic contribution, allowing us to separate the two components at various temperatures (see the Supplementary Information~\cite{SM} for details).

\begin{figure*}[ht]
\centering
\includegraphics[width=17cm]{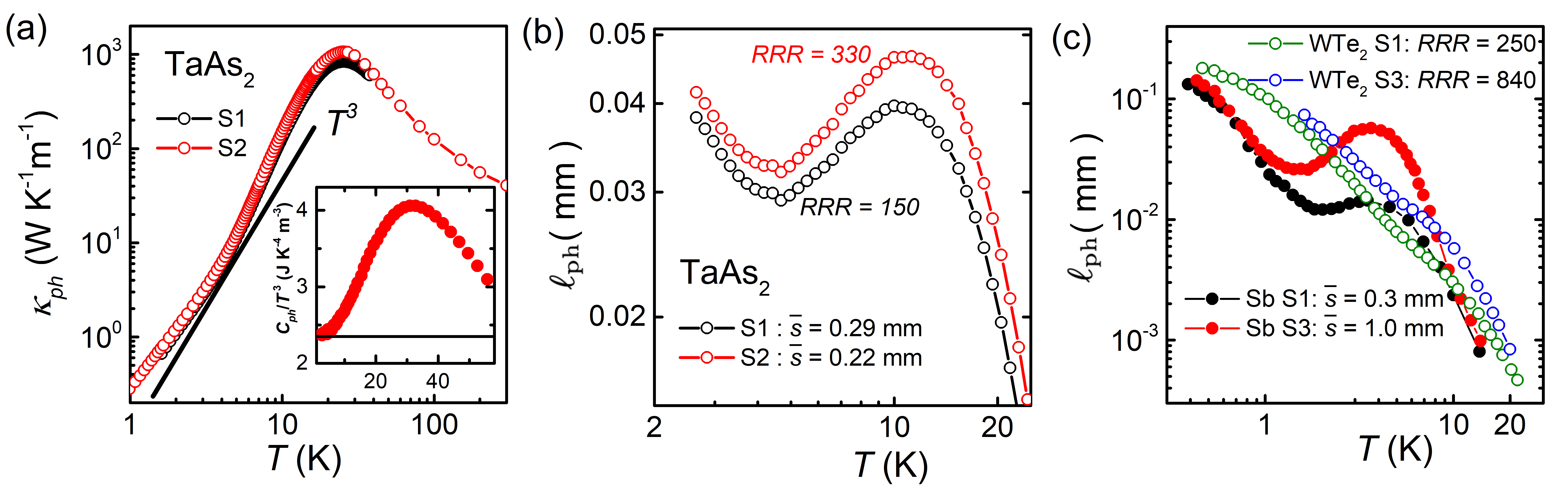}
\caption{\textbf{Phonon Poiseuille flow in TaAs$_2$ and other semimetals.}
(a) Phonon thermal conductivity $\kappa_{\mathrm{ph}}$ as a function of temperature. A faster-than-$T^3$ behavior is observed at intermediate temperatures. Inset: phonon heat capacity $C_{\mathrm{ph}}$ divided by $T^3$ as a function of temperature. The deviation of $C_{\mathrm{ph}}$ from $T^3$ dependence can be attributed to non-linear dispersion (see Supplementary Information\cite{SM}).
(b) Temperature-dependent phonon mean free path $\ell_{\mathrm{ph}} = 3\kappa_{\mathrm{ph}}/(C_{\mathrm{ph}} v_s)$, extracted from the data in (a). The effective sample size $\bar{s}$ is defined as $\sqrt{wt}$, where $w$ and $t$ are the sample width and thickness, respectively.
(c) Phonon mean free path of Sb and WTe$_2$ (data adapted from Refs.~\cite{Xie2024} and \cite{Jaoui2022}). In these materials, $\ell_{\mathrm{ph}}$ exhibits non-monotonic behavior, and its size- or impurity dependence is indicative of phonon hydrodynamics. However, $\ell_{\mathrm{ph}}$ remains far below the sample size in all cases, suggesting strong $e$-$ph$ coupling.}
\label{fig2}
\end{figure*}

\noindent\textbf{Phonon Poiseuille Flow.} 
The phononic thermal conductivity of both samples peaks at 25~K, with a faster-than-$T^3$ behavior between 5 and 15~K, as shown in Figure~\ref{fig2}(a). A slight difference between the two samples emerges near the peak but becomes more evident in their phonon mean free path $\ell_{\mathrm{ph}}$, defined as $\ell_{\mathrm{ph}} = 3\kappa_{\mathrm{ph}} / (C_{\mathrm{ph}} v_s)$. Here, $v_s$ is the average sound velocity, equal to $3678~\text{m·s}^{-1}$~\cite{Nobin2023}, and $C_{\mathrm{ph}}$ is the phonon heat capacity obtained by subtracting the $T$-linear electronic contribution from the total heat capacity (see Supplementary Information~\cite{SM} and inset of Figure~\ref{fig2}(a)). Although $C_{\mathrm{ph}}$ also deviates upward from $T^3$ behavior (similar deviations have been reported in Sb~\cite{Jaoui2022}), $\ell_{\mathrm{ph}}$, which is proportional to the ratio $\kappa_{\mathrm{ph}}/C_{\mathrm{ph}}$, still exhibits a non-monotonic temperature dependence with a Knudsen minimum reminiscent of the phonon Poiseuille flow predicted by Gurzhi~\cite{Gurzhi1968}.

The phonon Poiseuille flow regime occurs at intermediate temperatures, between the ballistic regime (where boundary scattering dominates) and the Ziman regime (where Umklapp scattering dominates)~\cite{Machida2024}. It satisfies $\ell^N \ll \bar{s} \ll \ell^U$, where $\bar{s}$ is the effective sample size, and $\ell^N$ and $\ell^U$ are the Normal and Umklapp scattering mean free paths, respectively. In this regime, the phonon mean free path scales as $\ell_{\mathrm{ph}} \sim \bar{s}^2 / \ell^N$, as previously reported in black phosphorus~\cite{Machida2018} and bismuth~\cite{kopylov1973}. In the semimetals TaAs$_2$, Sb~\cite{Jaoui2022}, and WTe$_2$~\cite{Xie2024}, the size- or purity-dependent non-monotonic $\ell_{\mathrm{ph}}$ shown in Figures~\ref{fig2}(c) and (d) is indicative of Poiseuille flow. However, as the temperature decreases well below the Poiseuille flow window, the ballistic regime is not reached, as evidenced by the fact that $\ell_{\mathrm{ph}}$ remains far below the typical sample size and becomes independent of both size and impurity. This anomaly in semimetals can be attributed to frequent $e$-$ph$ scattering, which reduces the phonon mean free path owing to strong $e$-$ph$ coupling.

\begin{figure*}[ht]
\centering
\includegraphics[width=17cm]{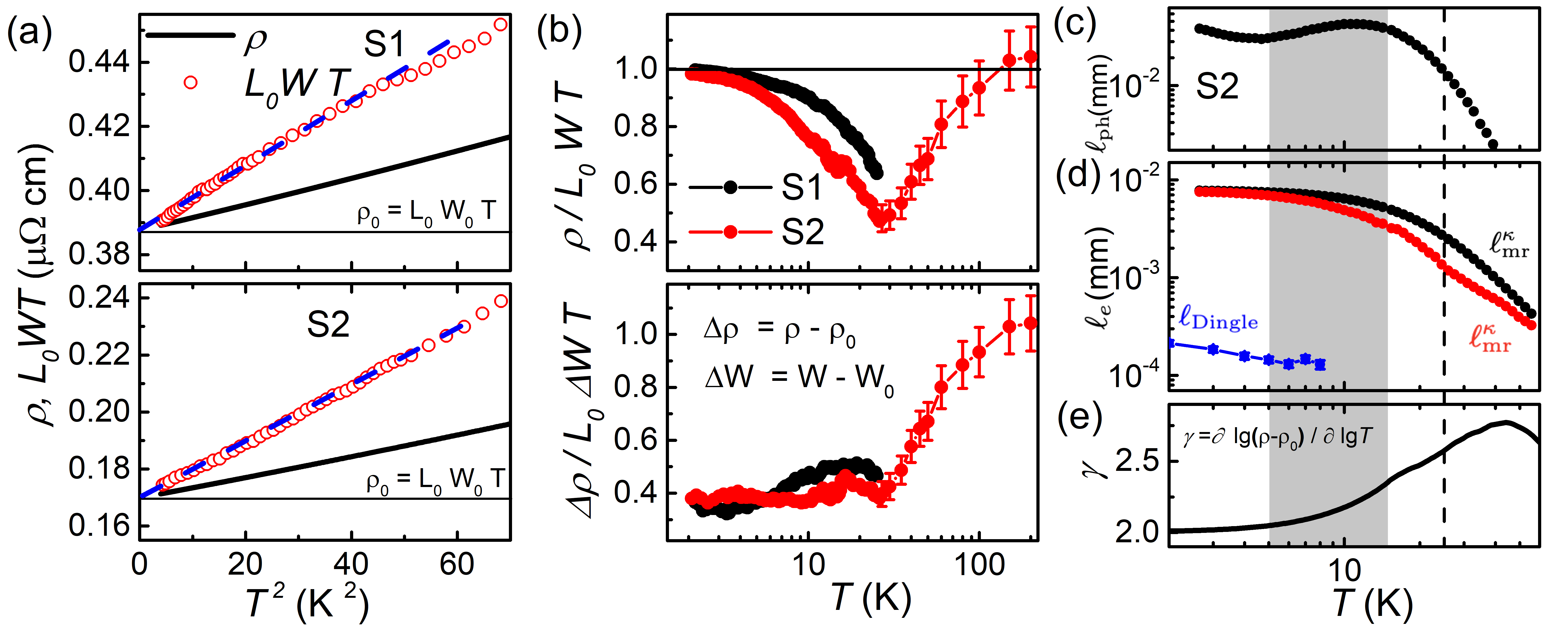}
\caption{\textbf{Electronic transport properties in TaAs$_2$.}
(a) Temperature dependence of the electrical and thermal resistivities of electrons in two samples with different impurity levels, plotted as $\rho$ versus $T^2$ and $L_0 W T$ versus $T^2$, where the thermal resistivity is defined as $W = 1/\kappa$.
(b) Upper panel: electronic Lorenz ratio $L/L_0 = \rho/(L_0 W T)$ as a function of temperature, with $L_0$ the Sommerfeld value. $L/L_0 = 1$ indicates recovery of the Wiedemann--Franz (WF) law. Lower panel: relative Lorenz ratio after subtracting the impurity contribution (i.e., the intercepts in panel (a)). The collapse of the data from the two samples indicates that $e$-$e$ scattering is responsible for the violation of the WF law below 20~K.
(c), (d), and (e) plot the phonon mean free path $\ell_{\mathrm{ph}}$, the electron mean free path $\ell_e$, and the temperature exponent of resistivity $\gamma$, respectively, as functions of temperature on the same temperature scale. The dashed line marks the $\kappa_{\mathrm{ph}}$ peak position. The gray shaded region indicates the phonon Poiseuille flow temperature window from 5 to 15~K. Within this range, $\gamma$ slightly exceeds 2, suggesting rare momentum-relaxing $e$-$ph$ scattering. In (d), the Dingle mean free path $\ell_{\mathrm{Dingle}}$, extracted from the quantum lifetime (see Supplementary Information~\cite{SM}), is nearly two orders of magnitude smaller than both the momentum-relaxing mean free path $\ell_{\mathrm{mr}}^{\sigma}$ (derived from electrical conductivity) and the energy-relaxing mean free path $\ell_{\mathrm{mr}}^{\kappa}$ (derived from electronic thermal conductivity). This indicates that frequent phonon-mediated $e$-$e$ scattering outweighs momentum-relaxing scattering, giving rise to electron hydrodynamics.}
\label{fig3}
\end{figure*}

\noindent\textbf{Electron Hydrodynamics.} 
We now turn to electronic transport. The temperature exponent $\gamma$ of the resistivity, defined by $\rho = \rho_0 + A_{\gamma}T^{\gamma}$, can be extracted using $\gamma = \partial \ln(\rho - \rho_0)/\partial \ln T$~\cite{Jaoui2022,Xie2024}. As shown in the inset of Figure~\ref{fig1}(b), $\gamma$ is similar for both TaAs$_2$ samples. Below 4~K, $\gamma$ is close to the Fermi liquid value of 2. Above 4~K, a Bloch-Gr\"uneisen regime emerges due to $e$-$ph$ scattering. In many metals~\cite{Jaoui2022}, $\gamma$ typically increases to a peak value of approximately 5 at intermediate temperatures owing to inelastic $e$-$ph$ scattering, and then saturates to unity at higher temperatures as $e$-$ph$ scattering becomes quasi-elastic. In TaAs$_2$, however, $\gamma$ reaches only 2.8, which is also the case in Sb~\cite{Jaoui2022} and WTe$_2$~\cite{Xie2024}, reflecting the coexistence of $e$-$e$ and $e$-$ph$ scattering at intermediate temperatures.

In the Fermi liquid regime, both the electrical resistivity $\rho$ and the thermal resistivity $L_0 W T$ (where $W = 1/\kappa_e$) follow a $T^2$ dependence, but the prefactor $B$ of $L_0 W T$ is larger than the prefactor $A$ of $\rho$, as shown in Figure~\ref{fig3}(a). The Fermi pockets of TaAs$_2$ are small and located near the boundary of the first Brillouin zone~\cite{Sun2020}, which still allows Umklapp $e$-$e$ scattering to occur as the origin of the $T^2$ resistivity.

The temperature dependence of the mismatch between electrical and electronic thermal resistivities, known as the violation of the Wiedemann-Franz (WF) law, is plotted in the upper panel of Figure~\ref{fig3}(b). The Lorenz ratio $L_e/L_0$ deviates from unity at intermediate temperatures but recovers to $\sim 1$ both at room temperature and in the zero-temperature limit. The recovery of the WF law at low and high temperatures originates from elastic impurity scattering and quasi-elastic $e$-$ph$ scattering, respectively. In the intermediate temperature range from 4 to 30~K, $L_e/L_0$ falls significantly below unity, with a sample-dependent magnitude. After subtracting the impurity contribution from both resistivities and normalizing their ratio by $L_0$, the resulting curve, shown in the lower panel of Figure~\ref{fig3}(b), becomes sample-independent and remains nearly flat at approximately 0.4, which equals the ratio of the prefactors of the $T^2$ thermal and electrical resistivities. Within this range, the violation of the WF law is dominated by $e$-$e$ scattering, with no evident contribution from $e$-$ph$ scattering.

In the phonon Poiseuille flow temperature window, the WF law violation and $\gamma \ll 5$, as shown in Figure~\ref{fig3} (e), indicate rare momentum-relaxing $e$-$ph$ scattering. However, owing to strong $e$-$ph$ coupling in the presence of phonon hydrodynamics, one can envisage a two-step process in which momentum from electrons is transferred to the phonon bath (where phonons form a fluid) and subsequently returned to the electron bath. This mechanism has recently been proposed as phonon-mediated $e$-$e$ scattering~\cite{Vool2021, Wang2022} and is believed to give rise to electron hydrodynamics. In parallel, the Dingle mean free path $\ell_{\text{Dingle}}$, extracted from the quantum lifetime (see Supplementary Information~\cite{SM}), is nearly two orders of magnitude smaller than both the transport mean free path $\ell_{\text{mr}}^{\sigma}$ (momentum-relaxing, from electrical conductivity) and $\ell_{\text{mr}}^{\kappa}$ (energy-relaxing, from thermal conductivity). This large difference indicates the existence of phonon-mediated $e$-$e$ scattering~\cite{Jaoui2022}, which conserves both momentum and energy of the electrons. When this process outweighs momentum-relaxing collisions, electron hydrodynamics emerges\cite{Wang2022}.

\begin{figure*}[ht]
\centering
\includegraphics[width=15.5cm]{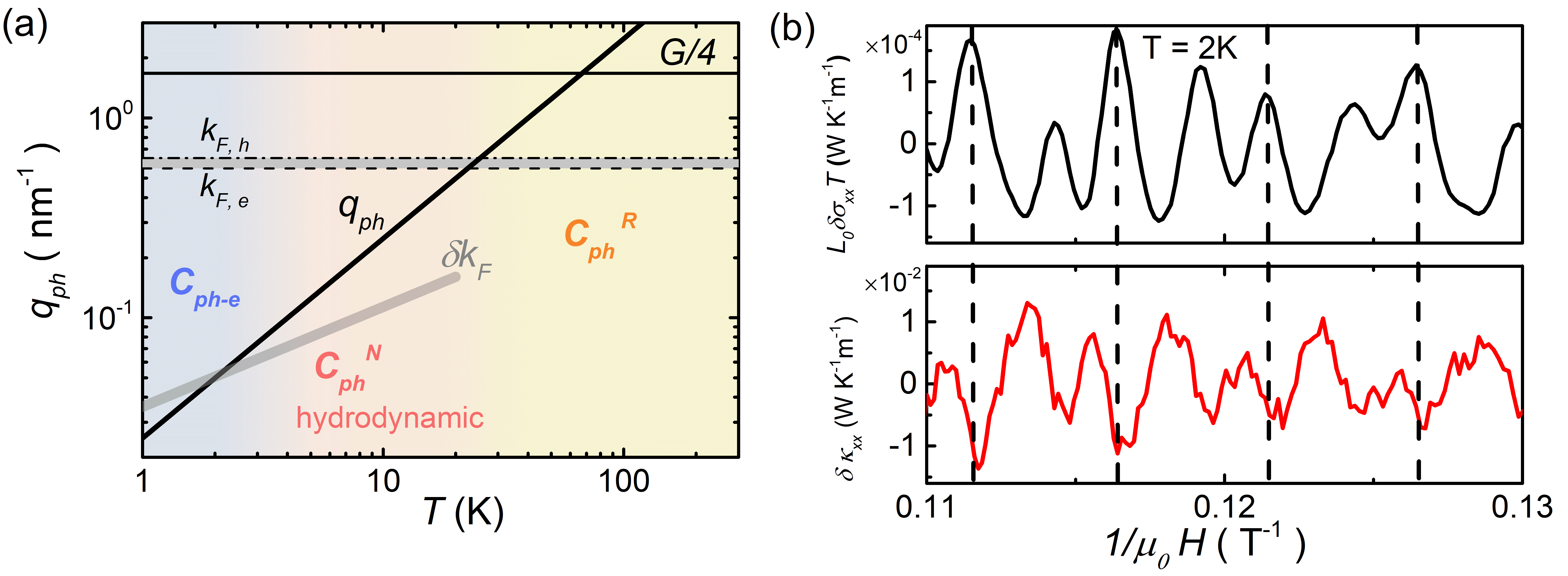}
\caption{\textbf{Evolution of wave vectors and quantum oscillations of lattice thermal conductivity in TaAs$_2$.}
(a) Comparison of the characteristic phonon wave vector $q_{\mathrm{ph}} \sim k_{\mathrm{B}}T/\hbar v_s$ with the average Fermi wave vector of $k_{\mathrm{F}}$. We note that $\kappa_{\mathrm{ph}}$ peaks when $q_{\mathrm{ph}}$ becomes comparable to $k_{\mathrm{F},e}$ and $k_{\mathrm{F},h}$, indicating significant $e$-$ph$ coupling. In the hydrodynamic regime, $q_{\mathrm{ph}}$ approaches the thermal smearing of the Fermi pockets, $\delta k_{\mathrm{F}} = \sqrt{2\pi m^* k_{\mathrm{B}} T}/\hbar$, which provides the phase space for phonon-mediated $e$-$e$ scattering.
(b) Comparison of quantum oscillations in the electronic ($L_0 \delta\sigma_{xx} T$) and total ($\delta\kappa_{xx}$) thermal conductivity. Since $\kappa_e$ is suppressed by the magnetic field, $\delta\kappa_{xx}$ can be safely attributed to phonons (see Supplementary Information~\cite{SM}). Its amplitude is two orders of magnitude larger than that of the electronic component and exhibits an anti-phase relation, both of which point to strong electron-phonon coupling.}
\label{fig4}
\end{figure*}

\begin{table*}[ht!]
\centering
\begin{tabular}
{m{1.5cm}<{\centering} m{1.8cm}<{\centering} m{1.8cm}<{\centering} m{1.8cm}<{\centering} m{1.3cm}<{\centering} m{1.0cm}<{\centering} m{1.0cm}<{\centering} m{1.2cm}<{\centering} m{2.0cm}<{\centering}  m{1.5cm}<{\centering}} 
\hline
Sample  & $n = p$  & $A$ & $B$ & $\bar{v}_s$  & $\lambda_{ph}$ & $\lambda_F$ & $\lambda_{ph}/\lambda_F$ & hydrodynamic & reference  \\
  &  (10$^{19}$ cm$^{-3}$) & ($n \Omega cm K^{-2}$) & ($n \Omega cm K^{-2}$) & (m S$^{-1}$) &  (nm) & (nm) &  &  \\
\hline
Bi & 0.03 & 12 & 35 &  1100 & 17.5 & 44.9 & 0.39 & $ph$ fluid & \cite{Gourgout2024,zhu2011,Boxus1981}\\
WP$_2$ & 250 & 0.017 & 0.074 & 3000 & 14.3  & 2.3 & 6.2 & $e$ fluid & \cite{Jaoui2018, Gooth2018}  \\
WTe$_2$ & 6.8 & 3.9 & 10 & 2200 & 6.4  & 7.8 & 0.8 & $e$-$ph$ bifluid & \cite{Xie2024, Zhu2015}  \\
TaAs$_2$ & 0.7 & 0.35 & 1 & 3678 & 7  & 9.1 & 0.77 & $e$-$ph$ bifluid & \cite{Nobin2023} \\
Sb & 5.5 & 0.3 & 0.6 & 2900 & 15.4  & 11.4 & 1.35 & $e$-$ph$ bifluid & \cite{Jaoui2022} \\
\hline
\end{tabular}
\caption{\textbf{Transport parameters of compensated semimetals.} 
$A$ and $B$ are the prefactors of the $T^2$ electrical and thermal resistivities, respectively. $\bar{v}_s$ is the average sound velocity, $\lambda_F$ is the average Fermi wavelength, and $\lambda_{\mathrm{ph}}$ is the phonon wavelength ($\lambda_{\mathrm{ph}} = 2\pi\hbar v_s / k_B T$) at the peak temperature of the lattice thermal conductivity. TaAs$_2$, Sb, and WTe$_2$ (to a lesser extent) exhibit $e$-$ph$ bifluid behavior. In these materials, $\lambda_{\mathrm{ph}}$ is close to $\lambda_F$, which facilitates the formation of the $e$-$ph$ bifluid. In Bi, the carrier density is too low, and in WP$_2$ it is too high; consequently, they exhibit only phonon hydrodynamics and electron hydrodynamics, respectively.}
\label{Table1}
\end{table*}

\section{DISCUSSION}

The aforementioned phononic thermal transport, together with electronic transport studies, all point to hydrodynamic features within the temperature window of 5 to 15~K. This is where $e$-$ph$ interaction plays a crucial role. Figure~\ref{fig4}(a) compares the characteristic phonon wave vector $q_{\mathrm{ph}} = k_B T / \hbar v_s$ with the Fermi wave vector $k_F$ and the Fermi thermal smearing $\delta k_F = \sqrt{2 \pi m^* k_B T} / \hbar$. On one hand, the peak in $\kappa_{\mathrm{ph}}$ occurs when $q_{\mathrm{ph}}$ becomes comparable to $k_F$. This trend is also observed in Sb~\cite{Jaoui2022}, implying strong $e$-$ph$ coupling. On the other hand, within the hydrodynamic temperature range, $q_{\mathrm{ph}}$ approaches $\delta k_F$, providing sufficient phase space for momentum-conserving phonon-mediated $e$-$e$ scattering. Furthermore, strong $e$-$ph$ coupling is also indicated by the high-field thermal conductivity oscillation $\delta \kappa_{xx}$, attributed to the lattice. Its amplitude is two orders of magnitude larger than that expected from the electronic component $L_0 \delta\sigma_{xx} T$, and it exhibits an out-of-phase relation, as shown in Figure~\ref{fig4}(b). Similar observations have been reported in other semimetals~\cite{Jaoui2022, Xie2024, Bermond2025}.

All these observations support an $e$-$ph$ bifluid picture in which phonons play a central role. On one hand, normal $ph$-$ph$ scattering is the dominant process, enabling phonons to behave collectively. On the other hand, strong $e$-$ph$ coupling provides an additional scattering channel for $e$-$e$ interactions via phonon exchange, which facilitates electron hydrodynamics while simultaneously coupling the electron and phonon fluids. These macroscopic hydrodynamic behaviors are intimately linked to the electronic band structure and phonon spectrum of TaAs$_2$. Recent theoretical work~\cite{Wang2022} attributes this hydrodynamic behavior to the $d$-orbital electrons and the low crystal symmetry of TaAs$_2$. Also, studies on graphene heterostructures~\cite{Sadeghi2023} indicate that electron–flexural phonon interactions even contribute to both classical electron hydrodynamics and quantum Cooper pairing. 

%According to experiments, TaAs$_2$, antimony  (Sb), and to a lesser extent, WTe$_2$ are semimetals, showing the clearest signatures of hybrid $e$-$ph$ hydrodynamics. Table~\ref{Table1} lists the properties of five semimetals. It is remarkable that those hosting thermodynamics happen to have a carrier density in the range of $\approx 10^{19}$cm$^{-3}$. This help $e$-$ph$ collision cross section, which is influenced by density-dependent $\lambda_F$ and temperature-dependent $\lambda_{ph}$ to become optimal at cryogenic temperatures.
Among the semimetals examined, TaAs$_2$, Sb~\cite{Jaoui2022}, and, to a lesser extent, WTe$_2$~\cite{Xie2024} exhibit the clearest signatures of hybrid $e$-ph hydrodynamics, in contrast to WP$_2$~\cite{Jaoui2018, Gooth2018} (which shows purely electronic hydrodynamics) and Bi~\cite{Boxus1981} (which exhibits phonon hydrodynamic behavior), with their transport parameters as summarized in Table~\ref{Table1}. We identify two conditions as prerequisites for the emergence of the $e$-$ph$ bifluid. First, within our temperature window of interest, the thermal phonon wavelength $\lambda_{\mathrm{ph}}$ (which scales with temperature) becomes comparable to the Fermi wavelength $\lambda_F$ (determined by the carrier density), i.e., $\lambda_{\mathrm{ph}}/\lambda_F \sim 1$, thereby realizing the optimal $e$-ph coupling regime. Second, the phonon dispersion exhibits significant nonlinearity, as evidenced by the deviation of the specific heat from the Debye approximation as shown in the inset of Figure~\ref{fig2}(a), which may signal enhanced normal $ph$-$ph$ collisions. The lesser extent of phonon Poiseuille flow in WTe$_2$ compared to Sb and TaAs$_2$ is evidenced by the non-monotonic phonon mean free path shown in Figure~\ref{fig2}(c), which exhibits a direct correlation with the phonon heat capacity $C_{\mathrm{ph}}$. In Sb and TaAs$_2$, $C_{\mathrm{ph}}/T^3$ is non-monotonic, while in WTe$_2$ it is not (see Fig.~S6 in the Supplementary Information~\cite{SM}). This implies that acoustic phonons have a significantly nonlinear dispersion, which amplifies the phase space for normal $ph$-$ph$ scattering, a condition essential for Poiseuille flow.
Although semimetals have attracted extensive attention primarily due to their topological properties, the hydrodynamic picture we propose here does not invoke any topological ingredients~\cite{link2018, Sun2018} and is therefore generic to semimetals. The coupled e-ph hydrodynamic has also been explored theoretically using semiclassical approaches~\cite{coulter2026}.

In summary, we have performed systematic electrical and thermal transport measurements on the compensated semimetal TaAs$_2$. Our results demonstrate that both phonons and charge carriers enter the hydrodynamic regime in the temperature range of 5 to 15~K, and that phonon-mediated $e$-$e$ scattering plays a central role in electron hydrodynamics. Optimal $e$-ph coupling happens when the phonon wavelength becomes comparable to the Fermi wavelength at intermediate temperatures, correlating the electron and phonon fluids. These findings pave the way for understanding the fundamental physics of hydrodynamic transport in multi-particle interacting systems.

\section{ACKNOWLEDGMENTS}

This work was supported by The
National Key Research and Development Program of
China (Grants No.2023YFA1609600,
No.2024YFA1611200 and No.2022YFA1403500), the
National Science Foundation of China (Grants
No.12304065, No.51821005, No.12004123,
No.51861135104 and No.11574097), the Fundamental
Research Funds for the Central Universities (Grant
No.2019kfyXMBZ071), the Hubei Provincial Natural
Science Foundation (2025AFA072).

\bibliography{TaAs2}

@article{Xie2024,
   author = {Xie, Wei and Yang, Feng and Xu, Liangcai and Li, Xiaokang and Zhu, Zengwei and Behnia, Kamran},
   title = {{Purity-dependent Lorenz number, electron hydrodynamics and electron-phonon coupling in WTe$_2$}},
   journal = {Science China Physics, Mechanics $\&$ Astronomy},
   volume = {67},
   number = {8},
   pages = {287014},
   ISSN = {1869-1927},
   DOI = {10.1007/s11433-024-2404-0},
   url = {https://doi.org/10.1007/s11433-024-2404-0},
   year = {2024},
   type = {Journal Article}
}

@article{Luo2016,
   author = {Luo, Yongkang and McDonald, R. D. and Rosa, P. F. S. and Scott, B. and Wakeham, N. and Ghimire, N. J. and Bauer, E. D. and Thompson, J. D. and Ronning, F.},
   title = {Anomalous electronic structure and magnetoresistance in {TaAs$_2$}},
   journal = {Scientific Reports},
   volume = {6},
   number = {1},
   pages = {27294},
   ISSN = {2045-2322},
   DOI = {10.1038/srep27294},
   url = {https://doi.org/10.1038/srep27294},
   year = {2016},
   type = {Journal Article}
}

@article{Liu2020,
   author = {Liu, Xiao-Lei and Wang, Hong-Yuan and Su, Hao and Yu, Zhen-Hai and Guo, Yan-Feng},
   title = {Nontrivial topological states in the tantalum dipnictides {TaX$_2$ (X=As, P)}},
   journal = {Tungsten},
   volume = {2},
   number = {3},
   pages = {251-260},
   ISSN = {2661-8036},
   DOI = {10.1007/s42864-020-00058-2},
   url = {https://doi.org/10.1007/s42864-020-00058-2},
   year = {2020},
   type = {Journal Article}
}

@article{Wu2016,
   author = {Wu, Desheng and Liao, Jian and Yi, Wei and Wang, Xia and Li, Peigang and Weng, Hongming and Shi, Youguo and Li, Yongqing and Luo, Jianlin and Dai, Xi and Fang, Zhong},
   title = {Giant semiclassical magnetoresistance in high mobility TaAs2 semimetal},
   journal = {Applied Physics Letters},
   volume = {108},
   number = {4},
   pages = {042105},
   ISSN = {0003-6951},
   DOI = {10.1063/1.4940924},
   url = {https://doi.org/10.1063/1.4940924},
   year = {2016},
   type = {Journal Article}
}

@article{Wadge2022,
   author = {Wadge, Ashutosh S. and Grabecki, Grzegorz and Autieri, Carmine and Kowalski, Bogdan J. and Iwanowski, Przemysław and Cuono, Giuseppe and Islam, M. F. and Canali, C. M. and Dybko, Krzysztof and Hruban, Andrzej and Łusakowski, Andrzej and Wojciechowski, Tomasz and Diduszko, Ryszard and Lynnyk, Artem and Olszowska, Natalia and Rosmus, Marcin and Kołodziej, J. and Wiśniewski, Andrzej},
   title = {{Electronic properties of {TaAs$_2$} topological semimetal investigated by transport and ARPES}},
   journal = {Journal of Physics: Condensed Matter},
   volume = {34},
   number = {12},
   pages = {125601},
   ISSN = {0953-8984},
   DOI = {10.1088/1361-648X/ac43fe},
   url = {https://dx.doi.org/10.1088/1361-648X/ac43fe},
   year = {2022},
   type = {Journal Article}
}

@article{Hu2025,
   author = {Hu, Haihua and Feng, Xiaolong and Pan, Yu and Hasse, Vicky and Wang, Honghui and He, Bin and Felser, Claudia},
   title = {Multipocket synergy towards high thermoelectric performance in topological semimetal {TaAs$_2$}},
   journal = {Nature Communications},
   volume = {16},
   number = {1},
   pages = {119},
   ISSN = {2041-1723},
   DOI = {10.1038/s41467-024-55490-6},
   url = {https://doi.org/10.1038/s41467-024-55490-6},
   year = {2025},
   type = {Journal Article}
}

@article{Bandurin2016,
   author = {Bandurin, D. A. and Torre, I. and Kumar, R. Krishna and Ben Shalom, M. and Tomadin, A. and Principi, A. and Auton, G. H. and Khestanova, E. and Novoselov, K. S. and Grigorieva, I. V. and Ponomarenko, L. A. and Geim, A. K. and Polini, M.},
   title = {Negative local resistance caused by viscous electron backflow in graphene},
   journal = {Science},
   volume = {351},
   number = {6277},
   pages = {1055-1058},
   DOI = {10.1126/science.aad0201},
   url = {https://doi.org/10.1126/science.aad0201},
   year = {2016},
   type = {Journal Article}
}

@article{Zhu2015,
  author = {Zhu, Zengwei and Lin, Xiao and Liu, Juan and Fauqu\'e, Beno\^{\i}t and Tao, Qian and Yang, Chongli and Shi, Youguo and Behnia, Kamran},
  title = {Quantum Oscillations, Thermoelectric Coefficients, and the Fermi Surface of Semimetallic {WTe$_2$}},
  journal = {Phys. Rev. Lett.},
  volume = {114},
  issue = {17},
  pages = {176601},
  numpages = {5},
  year = {2015},
  month = {Apr},
  publisher = {American Physical Society},
  doi = {10.1103/PhysRevLett.114.176601},
  url = {https://link.aps.org/doi/10.1103/PhysRevLett.114.176601}
}

@article{Jaoui2021,
   author = {Jaoui, Alexandre and Fauqué, Benoît and Behnia, Kamran},
   title = {Thermal resistivity and hydrodynamics of the degenerate electron fluid in antimony},
   journal = {Nature Communications},
   volume = {12},
   number = {1},
   pages = {195},
   ISSN = {2041-1723},
   DOI = {10.1038/s41467-020-20420-9},
   url = {https://doi.org/10.1038/s41467-020-20420-9},
   year = {2021},
   type = {Journal Article}
}

@article{Gooth2018,
   author = {Gooth, J. and Menges, F. and Kumar, N. and S\"{u}\ss{}, V. and Shekhar, C. and Sun, Y. and Drechsler, U. and Zierold, R. and Felser, C. and Gotsmann, B.},
   title = {Thermal and electrical signatures of a hydrodynamic electron fluid in tungsten diphosphide},
   journal = {Nature Communications},
   volume = {9},
   number = {1},
   pages = {4093},
   ISSN = {2041-1723},
   DOI = {10.1038/s41467-018-06688-y},
   url = {https://doi.org/10.1038/s41467-018-06688-y},
   year = {2018},
   type = {Journal Article}
}

@article{Jaoui2018,
   author = {Jaoui, Alexandre and Fauqué, Benoît and Rischau, Carl Willem and Subedi, Alaska and Fu, Chenguang and Gooth, Johannes and Kumar, Nitesh and Süß, Vicky and Maslov, Dmitrii L. and Felser, Claudia and Behnia, Kamran},
   title = {Departure from the Wiedemann–Franz law in {WP$_2$} driven by mismatch in T-square resistivity prefactors},
   journal = {npj Quantum Materials},
   volume = {3},
   number = {1},
   pages = {64},
   ISSN = {2397-4648},
   DOI = {10.1038/s41535-018-0136-x},
   url = {https://doi.org/10.1038/s41535-018-0136-x},
   year = {2018},
   type = {Journal Article}
}

@article{Jaoui2022,
  title = {Formation of an Electron-Phonon Bifluid in Bulk Antimony},
  author = {Jaoui, Alexandre and Gourgout, Adrien and Seyfarth, Gabriel and Subedi, Alaska and Lorenz, Thomas and Fauqu\'e, Beno\^{\i}t and Behnia, Kamran},
  journal = {Phys. Rev. X},
  volume = {12},
  issue = {3},
  pages = {031023},
  numpages = {9},
  year = {2022},
  month = {Aug},
  publisher = {American Physical Society},
  doi = {10.1103/PhysRevX.12.031023},
  url = {https://link.aps.org/doi/10.1103/PhysRevX.12.031023}
}

@article{Gurzhi1968,
   author = {Gurzhi, R. N.},
   title = {HYDRODYNAMIC EFFECTS IN SOLIDS AT LOW TEMPERATURE},
   journal = {Soviet Physics Uspekhi},
   volume = {11},
   number = {2},
   pages = {255},
   ISSN = {0038-5670},
   DOI = {10.1070/PU1968v011n02ABEH003815},
   url = {https://dx.doi.org/10.1070/PU1968v011n02ABEH003815},
   year = {1968},
   type = {Journal Article}
}

@article{Crossno2016,
 author = {Jesse Crossno  and Jing K. Shi  and Ke Wang  and Xiaomeng Liu  and Achim Harzheim  and Andrew Lucas  and Subir Sachdev  and Philip Kim  and Takashi Taniguchi  and Kenji Watanabe  and Thomas A. Ohki  and Kin Chung Fong },
 title = {Observation of the Dirac fluid and the breakdown of the Wiedemann-Franz law in graphene},
 journal = {Science},
 volume = {351},
 number = {6277},
 pages = {1058-1061},
 year = {2016},
 doi = {10.1126/science.aad0343},
 URL = {https://www.science.org/doi/abs/10.1126/science.aad0343},
}

@article{Martelli2018,
  author = {Martelli, Valentina and Jim\'enez, Julio Larrea and Continentino, Mucio and Baggio-Saitovitch, Elisa and Behnia, Kamran},
  title = {Thermal Transport and Phonon Hydrodynamics in Strontium Titanate},
  journal = {Phys. Rev. Lett.},
  volume = {120},
  issue = {12},
  pages = {125901},
  numpages = {6},
  year = {2018},
  month = {Mar},
  publisher = {American Physical Society},
  doi = {10.1103/PhysRevLett.120.125901},
  url = {https://link.aps.org/doi/10.1103/PhysRevLett.120.125901}
}

@article{Machida2018,
author = {Yo Machida  and Alaska Subedi  and Kazuto Akiba  and Atsushi Miyake  and Masashi Tokunaga  and Yuichi Akahama  and Koichi Izawa  and Kamran Behnia },
title = {Observation of Poiseuille flow of phonons in black phosphorus},
journal = {Science Advances},
volume = {4},
number = {6},
pages = {eaat3374},
year = {2018},
doi = {10.1126/sciadv.aat3374},
URL = {https://www.science.org/doi/abs/10.1126/sciadv.aat3374},
}

@article{Nobin2023,
   author = {Nobin, Md Nadim Mahamud and Khan, Mithun and Islam, Syed Saiful and Ali, Md Lokman},
   title = {Pressure-induced physical properties in topological semi-metal {TaM2 (M = As, Sb)}},
   journal = {RSC Advances},
   volume = {13},
   number = {32},
   pages = {22088-22100},
   DOI = {10.1039/D3RA03085G},
   url = {http://dx.doi.org/10.1039/D3RA03085G},
   year = {2023},
   type = {Journal Article}
}

@article{deJong1995,
   author = {de Jong, M. J. M. and Molenkamp, L. W.},
   title = {Hydrodynamic Electron Flow in High-Mobility Wires},
   journal = {Phys. Rev. B},
   volume = {51},
   pages = {13389},
   year = {1995},
   type = {Journal Article},
   doi = {10.1103/physrevb.51.13389}
}

@article{Beck1974,
   author = {Beck, H. and Meier, P. F. and Thellung, A.},
   title = {Phonon Hydrodynamics in Solids},
   journal = {Phys. Status Solidi A},
   volume = {24},
   pages = {11--62},
   year = {1974},
   type = {Journal Article},
   doi = {10.1002/pssa.2210240102}
}

@article{Machida2020,
   author = {Machida, Y. and Matsumoto, N. and Isono, T. and Behnia, K.},
   title = {Phonon Hydrodynamics and Ultrahigh-Room-Temperature Thermal Conductivity in Thin Graphite},
   journal = {Science},
   volume = {367},
   pages = {309--312},
   year = {2020},
   type = {Journal Article},
   doi = {10.1126/science.aaz8043}
}

@article{Coulter2019,
   author = {Coulter, J. and Osterhoudt, G. B. and Garcia, C. A. C. and Wang, Y. and Plisson, V. M. and Shen, B. and Ni, N. and Burch, K. S. and Narang, P.},
   title = {Uncovering Electron-Phonon Scattering and Phonon Dynamics in {Type-I} Weyl Semimetals},
   journal = {Phys. Rev. B},
   volume = {100},
   pages = {220301(R)},
   DOI = {10.1103/PhysRevB.100.220301},
   url = {https://link.aps.org/doi/10.1103/PhysRevB.100.220301},
   year = {2019},
   type = {Journal Article}
}

@article{Levchenko2020,
   author = {Levchenko, A. and Schmalian, J.},
   title = {Transport Properties of Strongly Coupled Electron-Phonon Liquids},
   journal = {Ann. Phys. (Amsterdam)},
   volume = {419},
   pages = {168218},
   DOI = {10.1016/j.aop.2020.168218},
   url = {https://doi.org/10.1016/j.aop.2020.168218},
   year = {2020},
   type = {Journal Article}
}

@article{Huang2021,
   author = {Huang, X. and Lucas, A.},
   title = {Electron-Phonon Hydrodynamics},
   journal = {Phys. Rev. B},
   volume = {103},
   pages = {155128},
   DOI = {10.1103/PhysRevB.103.155128},
   url = {https://link.aps.org/doi/10.1103/PhysRevB.103.155128},
   year = {2021},
   type = {Journal Article}
}

@article{Davison2014,
   author = {Davison, R. A. and Schalm, K. and Zaanen, J.},
   title = {Holographic Duality and the Resistivity of Strange Metals},
   journal = {Phys. Rev. B},
   volume = {89},
   pages = {245116},
   DOI = {10.1103/PhysRevB.89.245116},
   url = {https://link.aps.org/doi/10.1103/PhysRevB.89.245116},
   year = {2014},
   type = {Journal Article}
}

@article{Lucas2017,
   author = {Lucas, A. and Hartnoll, S. A.},
   title = {Resistivity Bound for Hydrodynamic Bad Metals},
   journal = {Proc. Natl. Acad. Sci. U.S.A.},
   volume = {114},
   number = {43},
   pages = {11344--11349},
   DOI = {10.1073/pnas.1711895114},
   url = {https://doi.org/10.1073/pnas.1711895114},
   year = {2017},
   type = {Journal Article}
}

@article{Vool2021,
   author = {Vool, U. and Hamo, A. and Varnavides, G. and Wang, Y. and Zhou, T. X. and Kumar, N. and Dovzhenko, Y. and Qiu, Z. and Garcia, C. A. C. and Pierce, A. T. and Gooth, J. and Anikeeva, P. and Felser, C. and Narang, P. and Yacoby, A.},
   title = {Imaging Phonon-Mediated Hydrodynamic Flow in {WTe$_2$}},
   journal = {Nat. Phys.},
   volume = {17},
   number = {10},
   pages = {1216--1222},
   DOI = {10.1038/s41567-021-01341-w},
   url = {https://doi.org/10.1038/s41567-021-01341-w},
   year = {2021},
   type = {Journal Article},
}

@article{Osterhoudt2021,
   author = {Osterhoudt, G. B. and Wang, Y. and Garcia, C. A. C. and Plisson, V. M. and Gooth, J. and Felser, C. and Narang, P. and Burch, K. S.},
   title = {Evidence for Dominant Phonon-Electron Scattering in Weyl Semimetal {WP$_2$}},
   journal = {Phys. Rev. X},
   volume = {11},
   number = {1},
   pages = {011017},
   DOI = {10.1103/PhysRevX.11.011017},
   url = {https://link.aps.org/doi/10.1103/PhysRevX.11.011017},
   year = {2021},
   type = {Journal Article}
}

@article{Kopylov1973,
   author = {Kopylov, V. and Mezhov-Deglin, L.},
   title = {Study of kinetic coefficients of bismuth at helium temperatures},
   journal = {Sov. Phys. JETP},
   volume = {65},
   number = {},
   pages = {720},
   DOI = {},
   url = {},
   year = {1973},
   type = {Journal Article}
}

@article{Bermond2025,
   author = {Bermond, B. and Wawrzyńczak, R. and Zherlitsyn, S. and Kotte, T. and Helm, T. and Gorbunov, D. and Gu, G. and Li, Q. and Janasz, F. and Meng, T. and Menges, F. and Felser, C. and Wosnitza, J. and Grushin, A. and Carpentier, D. and Gooth, J. and Gałeski, S.},
   title = {Giant quantum oscillations in thermal transport in low-density metals via electron absorption of phonons},
   journal = {Proc. Natl. Acad. Sci. U.S.A.},
   volume = {122},
   number = {10},
   pages = {e2408546122},
   DOI = {10.1073/pnas.2408546122},
   url = {https://doi.org/10.1073/pnas.2408546122},
   year = {2025},
   type = {Journal Article}
}

@article{Bhargava1967,
   author = {Bhargava, R. N.},
   title = {de Haas-van Alphen and Galvanomagnetic Effect in {Bi and Bi-Pb} Alloys},
   journal = {Phys. Rev.},
   volume = {156},
   number = {3},
   pages = {785--797},
   DOI = {10.1103/PhysRev.156.785},
   url = {https://link.aps.org/doi/10.1103/PhysRev.156.785},
   year = {1967},
   type = {Journal Article}
}

@article{Culbert1967,
   author = {Culbert, H. V.},
   title = {Low-Temperature Specific Heat of Arsenic and Antimony},
   journal = {Phys. Rev.},
   volume = {157},
   number = {3},
   pages = {560--563},
   DOI = {10.1103/PhysRev.157.560},
   url = {https://link.aps.org/doi/10.1103/PhysRev.157.560},
   year = {1967},
   type = {Journal Article}
}

@article{DiazSanchez2007,
   author = {Díaz-Sánchez, L. E. and Romero, A. H. and Cardona, M. and Kremer, R. K. and Gonze, X.},
   title = {Effect of the Spin-Orbit Interaction on the Thermodynamic Properties of Crystals: Specific Heat of Bismuth},
   journal = {Phys. Rev. Lett.},
   volume = {99},
   number = {16},
   pages = {165504},
   DOI = {10.1103/PhysRevLett.99.165504},
   url = {https://link.aps.org/doi/10.1103/PhysRevLett.99.165504},
   year = {2007},
   type = {Journal Article}
}

@book{Shoenberg2009,
   author = {Shoenberg, D.},
   title = {Magnetic Oscillations in Metals},
   publisher = {Cambridge University Press},
   year = {2009},
   type = {Book}
}

@article{Sun2018,
author = {Zhiyuan Sun  and Dmitry N. Basov  and Michael M. Fogler },
title = {Universal linear and nonlinear electrodynamics of a Dirac fluid},
journal = {Proceedings of the National Academy of Sciences},
volume = {115},
number = {13},
pages = {3285-3289},
year = {2018},
doi = {10.1073/pnas.1717010115},
URL = {https://www.pnas.org/doi/abs/10.1073/pnas.1717010115},
eprint = {https://www.pnas.org/doi/pdf/10.1073/pnas.1717010115},
}

@article{link2018,
  title = {Out-of-Bounds Hydrodynamics in Anisotropic Dirac Fluids},
  author = {Link, Julia M. and Narozhny, Boris N. and Kiselev, Egor I. and Schmalian, J\"org},
  journal = {Phys. Rev. Lett.},
  volume = {120},
  issue = {19},
  pages = {196801},
  numpages = {6},
  year = {2018},
  month = {May},
  publisher = {American Physical Society},
  doi = {10.1103/PhysRevLett.120.196801},
  url = {https://link.aps.org/doi/10.1103/PhysRevLett.120.196801}
}

@article{Gourgout2024,
   author = {Gourgout, A. and Marguerite, A. and Fauqué, B. and Behnia, K.},
   title = {Electronic thermal resistivity and quasiparticle collision cross section in semimetals},
   journal = {Phys. Rev. B},
   volume = {110},
   number = {15},
   pages = {155119},
   DOI = {10.1103/PhysRevB.110.155119},
   url = {https://link.aps.org/doi/10.1103/PhysRevB.110.155119},
   year = {2024},
   type = {Journal Article}
}

@article{krishna2017,
  title={Superballistic flow of viscous electron fluid through graphene constrictions},
  author={Krishna Kumar, R and Bandurin, DA and Pellegrino, FMD and Cao, Y and Principi, A and Guo, Haoyu and Auton, GH and Ben Shalom, M and Ponomarenko, Leonid Alexandrovich and Falkovich, G and others},
  journal={Nature Physics},
  volume={13},
  number={12},
  pages={1182--1185},
  year={2017},
  publisher={Nature Publishing Group UK London}
}

@article{Wang2022,
  title = {Generalized design principles for hydrodynamic electron transport in anisotropic metals},
  author = {Wang, Yaxian and Varnavides, Georgios and Sundararaman, Ravishankar and Anikeeva, Polina and Gooth, Johannes and Felser, Claudia and Narang, Prineha},
  journal = {Phys. Rev. Mater.},
  volume = {6},
  issue = {8},
  pages = {083802},
  numpages = {12},
  year = {2022},
  month = {Aug},
  publisher = {American Physical Society},
  doi = {10.1103/PhysRevMaterials.6.083802},
  url = {https://link.aps.org/doi/10.1103/PhysRevMaterials.6.083802}
}

@article{Sadeghi2023,
  title={Tunable electron--flexural phonon interaction in graphene heterostructures},
  author={Sadeghi, Mir Mohammad and Huang, Yajie and Lian, Chao and Giustino, Feliciano and Tutuc, Emanuel and MacDonald, Allan H and Taniguchi, Takashi and Watanabe, Kenji and Shi, Li},
  journal={Nature},
  volume={617},
  number={7960},
  pages={282--286},
  year={2023},
  publisher={Nature Publishing Group UK London}
}

@article{lifshitz1956,
  title={Theory of magnetic susceptibility in metals at low temperatures},
  author={Lifshitz, IM and Kosevich, Am M},
  journal={Sov. Phys. JETP},
  volume={2},
  number={4},
  pages={636--645},
  year={1956}
}

@article{behnia2022,
  title={On the origin and the amplitude of T-square resistivity in fermi liquids},
  author={Behnia, Kamran},
  journal={Annalen der Physik},
  volume={534},
  number={5},
  pages={2100588},
  year={2022},
  publisher={Wiley Online Library}
}

@article{kawabata2025,
  title = {Phonon hydrodynamic regimes in sapphire},
  author = {Kawabata, Takuya and Shimura, Kosuke and Ishii, Yuto and Koike, Minatsu and Yoshida, Kentaro and Yonehara, Shu and Yokoi, Kohei and Subedi, Alaska and Behnia, Kamran and Machida, Yo},
  journal = {Phys. Rev. Res.},
  volume = {7},
  issue = {3},
  pages = {033017},
  numpages = {9},
  year = {2025},
  month = {Jul},
  publisher = {American Physical Society},
  doi = {10.1103/ptds-chrz},
  url = {https://link.aps.org/doi/10.1103/ptds-chrz}
}

@article{guo2026hydrodynamicsviscouselectronfluid,
  title = {Hydrodynamics of the viscous electron fluid in cadmium},
  author = {Guo, Xiaodong and Li, Xiaokang and Fauqué, Benoît and Subedi, Alaska and Zhao, Lingxiao and Zhu, Zengwei and Behnia, Kamran},
  journal = {Phys. Rev. Lett.},
  pages = {},
  year = {2026},
  month = {Aug},
  publisher = {American Physical Society},
  doi = {10.1103/g32f-j6q2},
  url = {https://link.aps.org/doi/10.1103/g32f-j6q2}
}

@misc{coulter2026,
      title={Coupled electron-phonon hydrodynamics and viscous thermoelectric equations}, 
      author={Jennifer Coulter and Bogdan Rajkov and Michele Simoncelli},
      year={2026},
      eprint={2503.07560},
      archivePrefix={arXiv},
      primaryClass={cond-mat.mes-hall},
      url={https://arxiv.org/abs/2503.07560}, 
}

@misc{SM,
    howpublished = {See {Supplemental Material} for more details},
    	year = {2026},
}

@article{White01041958,
author = {G. K. White and S. B. Woods},
title = {The thermal and electrical resistivity of bismuth and antimony at low temperatures},
journal = {The Philosophical Magazine: A Journal of Theoretical Experimental and Applied Physics},
volume = {3},
number = {28},
pages = {342--359},
year = {1958},
publisher = {Taylor \& Francis},
doi = {10.1080/14786435808236822},
}

@article{Sun2020,
  title = {Topological metals induced by the Zeeman effect},
  author = {Sun, Song and Song, Zhida and Weng, Hongming and Dai, Xi},
  journal = {Phys. Rev. B},
  volume = {101},
  issue = {12},
  pages = {125118},
  numpages = {10},
  year = {2020},
  month = {Mar},
  publisher = {American Physical Society},
  doi = {10.1103/PhysRevB.101.125118},
  url = {https://link.aps.org/doi/10.1103/PhysRevB.101.125118}
}

@article{Zhu2011,
  title = {Angle-resolved Landau spectrum of electrons and holes in bismuth},
  author = {Zhu, Zengwei and Fauqu\'e, Beno\^{\i}t and Fuseya, Yuki and Behnia, Kamran},
  journal = {Phys. Rev. B},
  volume = {84},
  issue = {11},
  pages = {115137},
  numpages = {12},
  year = {2011},
  month = {Sep},
  publisher = {American Physical Society},
  doi = {10.1103/PhysRevB.84.115137},
  url = {https://link.aps.org/doi/10.1103/PhysRevB.84.115137}
}

@article{Boxus1981,
  title = {Size dependence of the transport properties of trigonal bismuth},
  author = {Boxus, J. and Uher, C. and Heremans, J. and Issi, J -P.},
  journal = {Phys. Rev. B},
  volume = {23},
  issue = {2},
  pages = {449--452},
  numpages = {0},
  year = {1981},
  month = {Jan},
  publisher = {American Physical Society},
  doi = {10.1103/PhysRevB.23.449},
  url = {https://link.aps.org/doi/10.1103/PhysRevB.23.449}
}

@article{Machida2024,
    author = {Machida, Yo and Martelli, Valentina and Jaoui, Alexandre and Fauqué, Benoît and Behnia, Kamran},
    title = {Phonon hydrodynamics in bulk insulators and semimetals},
    journal = {Low Temperature Physics},
    volume = {50},
    number = {7},
    pages = {574-583},
    year = {2024},
    month = {07},
    issn = {1063-777X},
    doi = {10.1063/10.0026323},
    url = {https://doi.org/10.1063/10.0026323},
}

\clearpage
\onecolumngrid

\begin{center}{\large\bf Supplemental Material for 'Electron-phonon coupled hydrodynamics in semimetal TaAs$_2$'}
\end{center}
\renewcommand{\thesection}{S\arabic{section}}
\renewcommand{\thetable}{S\arabic{table}}
\renewcommand{\thefigure}{S\arabic{figure}}
\renewcommand{\theequation}{S\arabic{equation}}

\setcounter{section}{0}
\setcounter{figure}{0}
\setcounter{table}{0}
\setcounter{equation}{0}

\section{Samples and Methods}

First, polycrystalline TaAs$_2$ was synthesized via a solid-state reaction by heating stoichiometric amounts of niobium powder (99.95\%) and tantalum powder (99.99\%) at 850$^{\circ}$C. Subsequently, an appropriate amount of iodine was added to the resulting polycrystalline powder as a transport agent. The cold and hot ends of the quartz tube were maintained at 850$^{\circ}$C and 950$^{\circ}$C, respectively. After reacting for 7 days, the tube was cooled naturally to room temperature, yielding bulk TaAs$_2$ single crystals. Due to the tendency of arsenic to form vacancies during crystal growth, single crystals grown by this method typically exhibit residual resistivity ratios ($RRR = \rho_{300\mathrm{K}}/\rho_{2\mathrm{K}}$) below 200. To obtain higher-quality single crystals, we replaced the tantalum powder with tantalum foil. During crystal growth, nucleation and growth occur directly on the tantalum foil, reducing impurity-related defects. Single crystals grown via this method generally achieve $RRR$ values exceeding 300. The obtained single crystals were then characterized for structure and composition using X-ray diffraction (XRD) and energy-dispersive X-ray spectroscopy (EDS). Table~\ref{TableS1} presents detailed information on the two studied samples, which are purer than those in previous studies~\cite{Luo2016, Hu2025}, with higher $RRR$ values exceeding 100.

TaAs$_2$ has a monoclinic structure with $\beta$ around $120^\circ$~\cite{Luo2016}. Heat or electric current is applied along the $b$-axis with the magnetic field parallel to the $c$-axis, as shown in the main text Figure~\ref{fig1}(c). The transport measurements were performed with home-built one-heater-two-thermometers or one-heater-two-thermocouples setups, which allowed us to measure the electrical resistivity and the thermal conductivity with the same electrodes. Above 10~K, we performed them in a physical property measurement system (Quantum Design) and between 1~K and 15~K, in a Leiden dilution refrigerator (CF-CS81-1700-Maglev). Between 1~K and 15~K, CX-1030 Cernox chips were used to measure the temperature, and type E thermocouples were used between 10~K and room temperature. The thermal gradient in the sample was produced through a 10~k$\Omega$ chip resistor. Electrical transport measurements were performed using a standard four-probe method.

\begin{table*}[ht!]
\centering
\begin{tabular}
{m{1.2cm}<{\centering} m{2.5cm}<{\centering} m{1.8cm}<{\centering} m{2.5cm}<{\centering} m{2.5cm}<{\centering}} 
\hline
TaAs$_2$  & Size($mm^3$)  & $\rho_0 (\mu \Omega cm)$ & $RRR=\frac{\rho_{300K}}{\rho_{2K}}$ & $\bar{s}=\sqrt{wt} (mm)$ \\ 
\hline
S1 & $1.6\times0.6\times0.14$ & 0.39 & 150 & 0.29 \\
\hline
S2 & $1.1\times0.5\times0.10$ & 0.17 & 330 & 0.22  \\
\hline
\end{tabular}
\caption{\textbf{Details of the samples}}
\label{TableS1}
\end{table*}

\section{Magneto-electric transport properties of TaAs$_2$}

\begin{table*}[ht!]
\centering
\begin{tabular}
{m{2cm}<{\centering} m{3.0cm}<{\centering} m{2.0cm}<{\centering} m{2.0cm}<{\centering} m{2.0cm}<{\centering} m{2.0cm}<{\centering}} \hline
$F(T)$  & $A_F(10^{-4}\AA^{-2})$  & $k_F(\AA^{-1})$ & $m^*(m_e)$  & $v_F(10^5 m/s)$  & $E_F(meV)$    \\
\hline
104 & 99 & 0.056 & 0.183 & 3.54 & 65  \\
131 & 125 & 0.063 & 0.179 & 4.07 & 84  \\
197 & 188 & 0.077 & 0.166 & 5.37 & 136  \\
218 & 208 & 0.081 & 0.172 & 5.45 & 145  \\
\hline
\end{tabular}
\caption{\textbf{The Fermi surface parameter}}
\label{TableS2}
\end{table*} 

\begin{figure*}[ht]
\centering
\includegraphics[width=17.5cm]{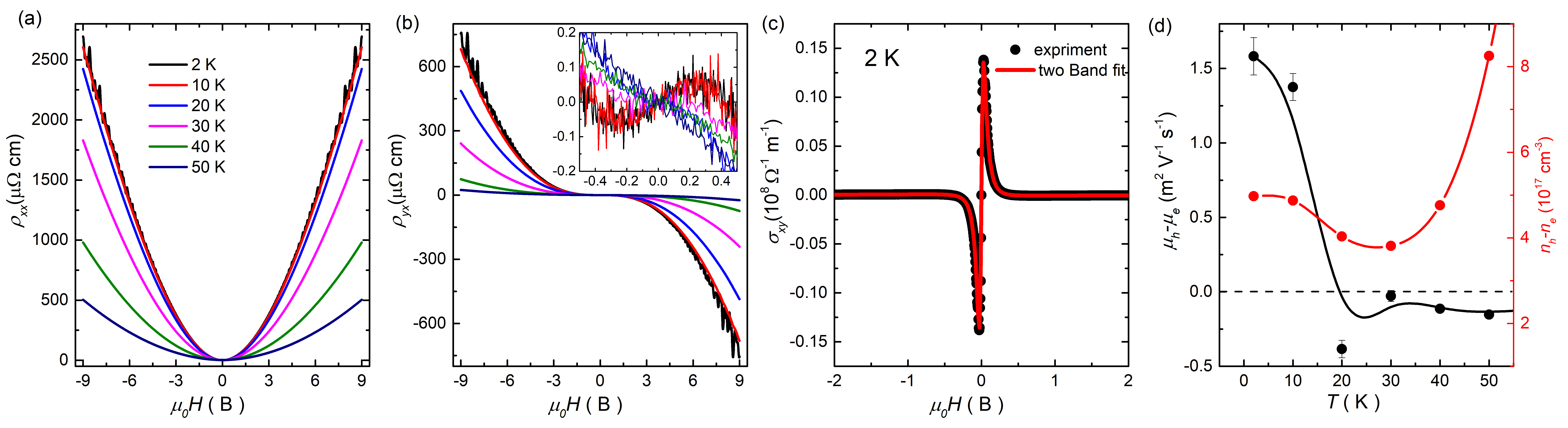}
\caption{\textbf{Magnetoresistivity and Hall effect in TaAs$_2$.}
(a) and (b) Magnetoresistivity $\rho_{xx}$ and Hall resistivity $\rho_{xy}$ as functions of magnetic field at various temperatures.
(c) Temperature dependence of the electron and hole mobility difference $\mu_h - \mu_e$ and the carrier concentration difference $n_h - n_e$. Both $\mu_h - \mu_e$ and $n_h - n_e$ are obtained by fitting $\sigma_{xy}$ in the low-field and high-field limits.
(d) Hall conductivity $\sigma_{xy}$ compared with the two-band model fitting. }
\label{figs1}
\end{figure*}

Figure~\ref{figs1} shows the magnetoresistivity $\rho_{xx}$ and the Hall resistivity $\rho_{yx}$ of TaAs$_2$. The quadratic field dependence of $\rho_{xx}$ and the nonlinear behavior of $\rho_{yx}$ indicate compensated charge carrier properties, consistent with previous experimental and density functional theory (DFT) calculation studies~\cite{Luo2016, Wu2016, Hu2025}. Using the two-band model for the Hall conductivity,
\begin{equation}
  \sigma_{xy}=\frac{\rho_{yx}}{\rho_{xx}^2+\rho_{yx}^2}=\frac{\sigma_h \mu_h B}{1+(\mu_h B)^2}-\frac{\sigma_e \mu_e B}{1+(\mu_e B)^2},
\end{equation}
where $\sigma_e$ and $\sigma_h$ are the zero-field conductivities of electrons and holes, respectively, we extracted the carrier densities $n$ and mobilities $\mu$ for electrons and holes, yielding values of $7\times 10^{18}~\text{cm}^{-3}$ and $10^5~\text{cm}^2~\text{V}^{-1}~\text{s}^{-1}$, respectively. Since $\mu$ in many materials is field-dependent, we employed approximations to examine the difference between electron and hole carriers. In the low-field limit ($\mu B \ll 1$) and assuming $\sigma_e = \sigma_h$, the Hall conductivity can be expressed as
\begin{equation}
  \sigma_{xy} \approx \sigma_e (\mu_h-\mu_e)B,
\end{equation}
while in the high-field limit ($\mu B \gg 1$),
\begin{equation}
  \sigma_{xy} \approx e^2(n_h-n_e)/B.
\end{equation}
We obtained the mobility difference $\mu_h-\mu_e$ at low field and the carrier density difference $n_h-n_e$, as shown in Figure~\ref{figs1}(c). A difference of less than 10\% between electron and hole carrier densities indicates well-compensated properties, similar to those observed in WTe$_2$~\cite{Zhu2015} and Bi~\cite{Bhargava1967}.

\begin{figure*}[ht]
\centering
\includegraphics[width=16cm]{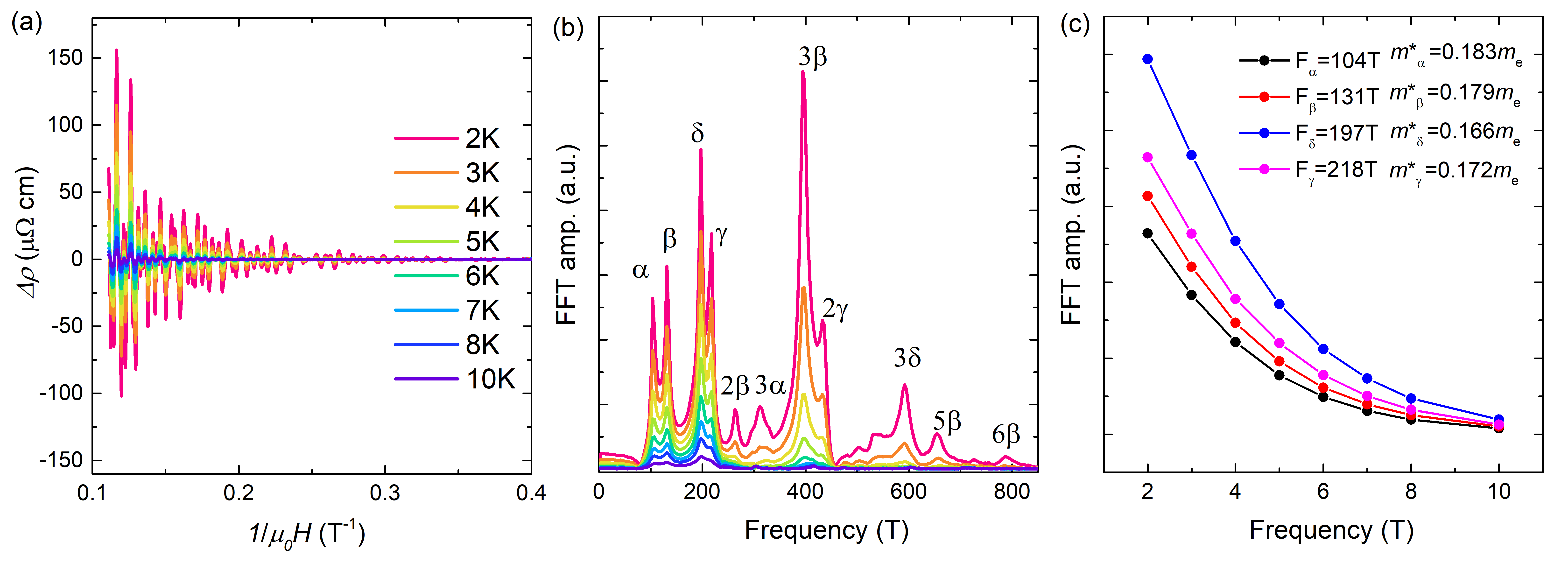}
\caption{\textbf{Quantum oscillation in TaAs$_2$.}
(a) $\rho_{xx}$ oscillation after subtracting a monotonic background, plotted as a function of inverse magnetic field.
(b) Fast Fourier transform (FFT) spectra of the $\rho_{xx}$ oscillations. $\alpha$, $\beta$, $\theta$, and $\gamma$ correspond to four extremal cross-sectional areas of the Fermi surface.
(c) Effective mass calculations for the four frequencies obtained from the Lifshitz–Kosevich formula\cite{Shoenberg2009}.}
\label{figs2}
\end{figure*}

As shown in Figure~\ref{figs1}(a), $\rho_{xx}$ at 2~K exhibits pronounced quantum oscillations. Figure~\ref{figs2} presents the fermiology analysis derived from these quantum oscillations. Using the Lifshitz-Kosevich (LK) formula~\cite{lifshitz1956},
\begin{equation}
  \Delta \rho_{xx} \propto R_T R_D \cos\left[2\pi\left(\frac{F}{B}-\gamma\right)+\Phi_D\right],
\end{equation}
where $R_T = (\alpha m^*T/B)/\sinh(\alpha m^*T/B)$ is the temperature attenuation factor and $R_D = \exp(-\alpha m^*T_D/B)$ is the Dingle attenuation factor, with $\alpha = 2\pi^2 k_B/\hbar e$. Using the Fast Fourier transform, we identified four quantum oscillation frequencies corresponding to four orbitals. By fitting $R_D$ and $R_T$, we obtained the effective masses of the four orbitals and the Dingle temperature $T_D$. Since $T_D = \hbar / 2\pi k_B \tau_q$, we extracted $\tau_q$, the quantum scattering time, which was used to estimate the Dingle scattering mean free path $\ell_q = v_F \tau_q$ presented in the main text.

From the fermiology analysis, we identified four frequencies with small effective masses, corresponding to complex electron and hole Fermi surfaces. The $\alpha$ and $\beta$ orbitals originate from electron and hole pockets, consistent with previous reports~\cite{Luo2016}. In contrast, the $\delta$ and $\gamma$ bands obtained in this study are inconsistent with previous work, where they were interpreted as the double frequency of $\alpha$ and its nearby shoulder frequency. This discrepancy may be attributed to differences in sample quality, particularly regarding arsenic (As) dopants. Notably, our sample exhibits a residual resistivity ratio ($RRR$) as high as 300, significantly larger than the value of 100 reported in previous studies~\cite{Luo2016}. The $\delta$ and $\gamma$ bands thus warrant careful further investigation. Table S1 summarizes the parameters of these four oscillation frequencies. In the main text, the Fermi velocity and effective mass correspond to the average values across the four bands.

Table 1 of the main text compares the transport properties of three compensated semimetals—Sb~\cite{Jaoui2018}, WTe$_2$~\cite{Xie2024}, and TaAs$_2$ (this work)—in which phonon-mediated electron hydrodynamics has been reported. Among them, TaAs$_2$ exhibits the highest phononic thermal conductivity and Debye temperature, both of which are favorable for the emergence of $e$-$ph$ hydrodynamics.

\section{Separating electronic and phononic thermal conductivity}

Figure~\ref{fig3} displays the thermal conductivity $\kappa$ as a function of the magnetic field. Below 27~K, $\kappa$ saturates at 1~T, leaving only the phononic thermal conductivity due to the high mobility of the charge carriers. Above 27~K, $\kappa$ consists of two components: a field-independent phonon term $\kappa_{\mathrm{ph}}$ and an electronic term $\kappa_e = \frac{T}{a + b B^n}$. We separate $\kappa_e$ and $\kappa_{\mathrm{ph}}$ by fitting the high-temperature, field-dependent $\kappa$ to this expression. This method has also been used in previous studies of WTe$_2$~\cite{Xie2024} and Sb~\cite{Jaoui2018}.

After separating $\kappa_e$, we obtained both the $T^2$-dependent resistivity and thermal resistivity. Figure~\ref{figs4} shows the magnitude of the $T^2$-dependent resistivity and thermal resistivity prefactors $A$ and $B$ for various semimetals as a function of their Fermi energy. Both $A$ and $B$ scale with the $e$-$e$ scattering phase space, $\propto (T / T_F)$. For a more detailed discussion, see Reference~\cite{behnia2022}.

\section{Phonon heat capacity and anharmonicity}

Figure~\ref{figs5} shows the specific heat $C$ of TaAs$_2$ down to 2~K. By fitting the plot of $C/T$ as a function of $T^2$, we separated the electronic component $C_e = \gamma T$ and the phononic component $C_{\mathrm{ph}} = \alpha T^3 + \beta T^5$~\cite{Culbert1967}. The deviation of $C_{\mathrm{ph}}$ from Debye theory may imply phonon anharmonicity at high temperatures, which has also been reported in As~\cite{Culbert1967}, Sb~\cite{Jaoui2022}, and Bi~\cite{DiazSanchez2007}. This effect is more clearly illustrated in the plot of $C_{\mathrm{ph}}/T^3$. Strong phonon anharmonicity in TaAs$_2$ and Sb may lead to frequent normal $ph$-$ph$ collisions, which in turn gives rise to phonon hydrodynamic transport, as shown in Figure~\ref{figs6}. The Debye temperature in the main text was obtained using $T_D = \sqrt[3]{\frac{12\pi^4 R}{5\alpha}} = \sqrt[3]{\frac{12 \times (3.1416)^4 \times 8.314}{5 \times 0.0831}} = 286~\text{K}$.

\begin{figure*}[ht]
\centering
\includegraphics[width=14cm]{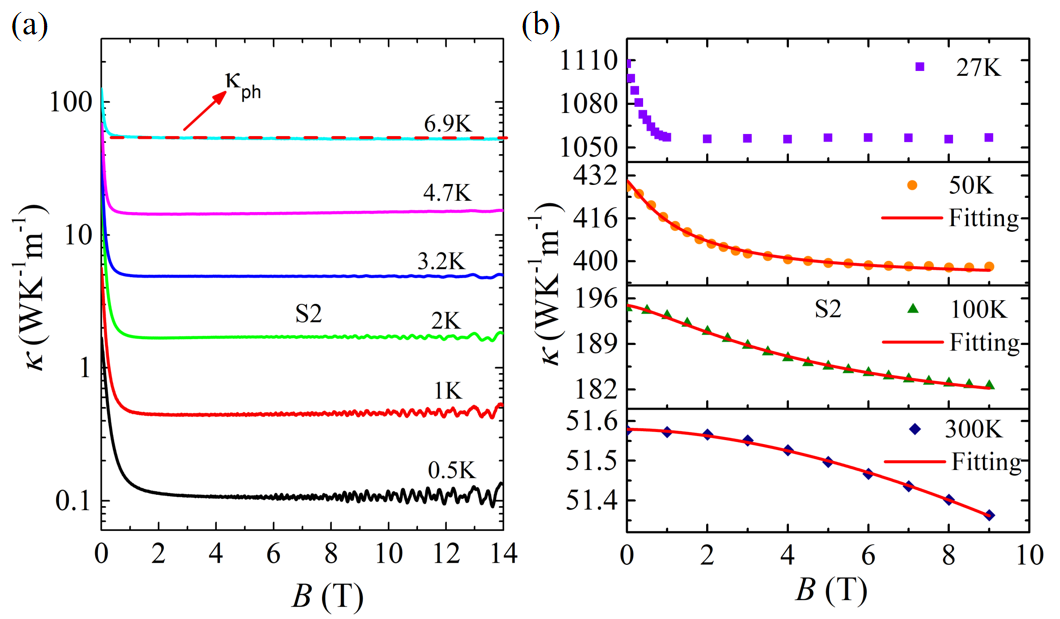}
\caption{\textbf{Subtracting electron and phonon thermal conductivity in magnetic field.}
(a) and (b) display the field-dependent thermal conductivity $\kappa$ as a function of temperature. Below 27~K, a magnetic field of 3~T suppresses the electronic thermal conductivity $\kappa_e$, leaving the phononic thermal conductivity $\kappa_{\text{ph}}$. Above 27~K, the field suppression becomes less pronounced, and the phononic thermal conductivity is then evaluated from a fit to $\kappa = \kappa_{\text{ph}} + T/(a + bB^n)$, shown as solid lines.}
\label{figs3}
\end{figure*}

\begin{figure*}[ht]
\centering
\includegraphics[width=15cm]{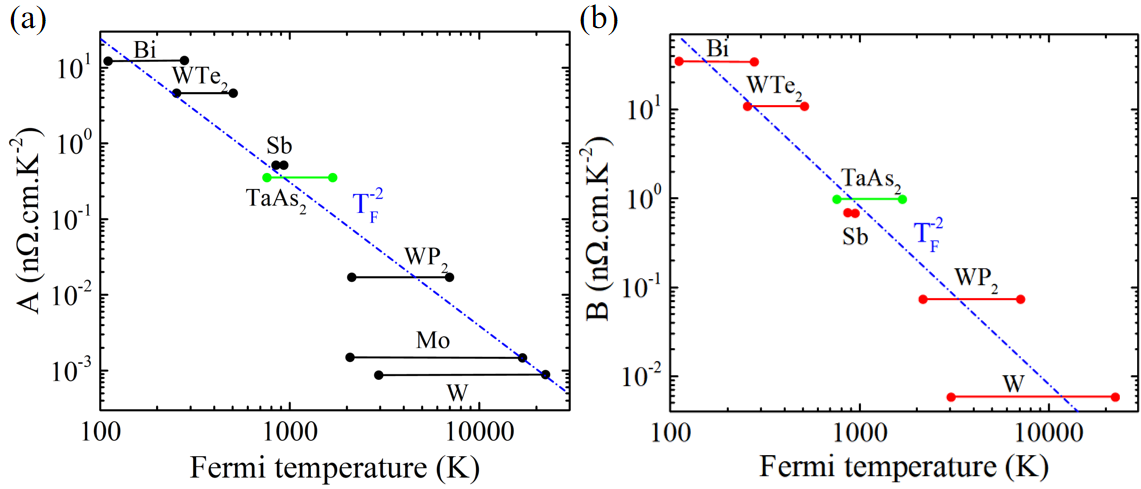}
\caption{\textbf{$T^2$ resistivity ($\rho$) and thermal resistivity $WT$ in semimetals.}
(a) and (b) display the prefactors of $T^2$ resistivity $\rho$ and thermal resistivity $WT$ in Bi, WTe$_2$, Sb, WP$_2$, Mo, W (summarized in \cite{Gourgout2024}), TaAs$_2$ (this work) as functions of the Fermi temperature $T_{\mathrm{F}}$ of electrons and holes. Note that Sb and TaAs$_2$ have similar carrier densities and effective masses.}
\label{figs4}
\end{figure*}

\begin{figure*}[ht]
\centering
\includegraphics[width=15cm]{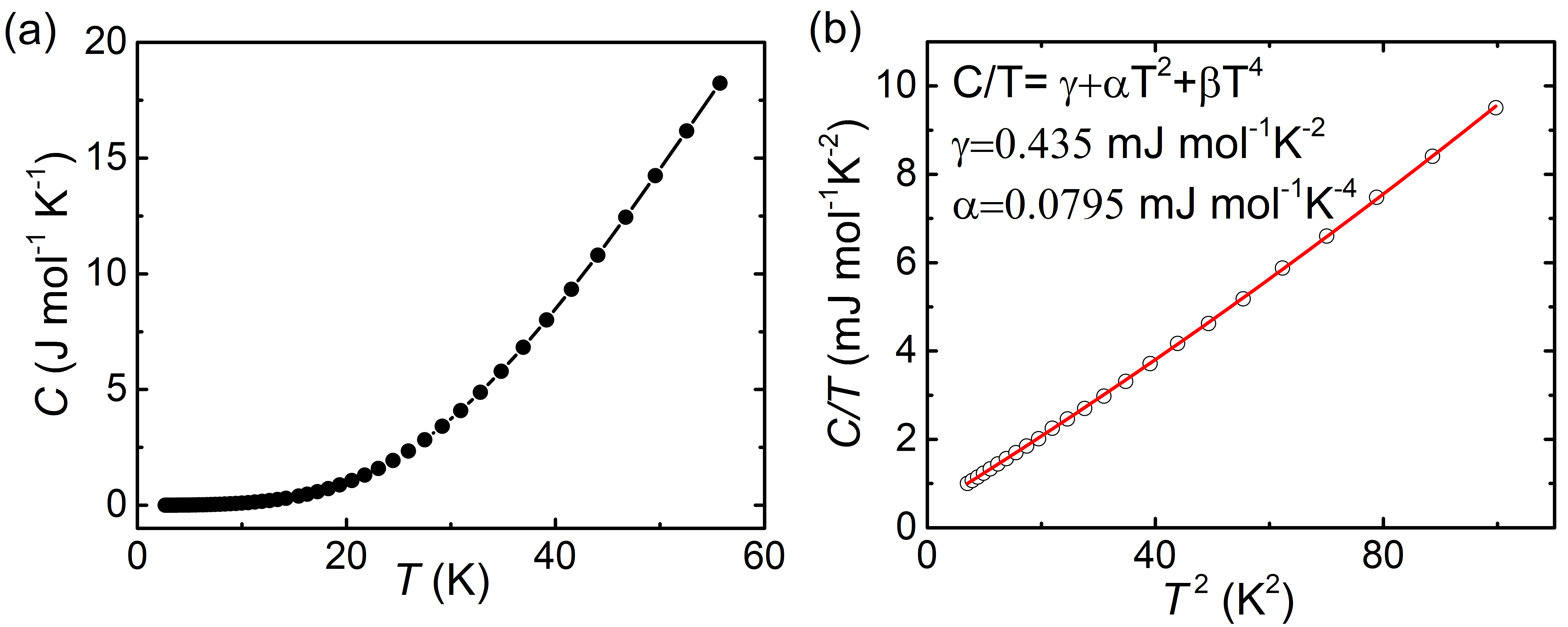}
\caption{\textbf{Heat capacity $C$ of TaAs$_2$.}
(a) $C$ as a function of temperature. (b) $C/T$ plotted as a function of temperature squared. The red line is a fit to $C/T = \gamma + \alpha T^2 + \beta T^4$, where the $\gamma$ and $\alpha$ terms correspond to the electronic and phonon heat capacity, respectively. The higher-order $\beta$ term accounts for phonon anharmonicity.}
\label{figs5}
\end{figure*}

\begin{figure*}[ht]
\centering
\includegraphics[width=8cm]{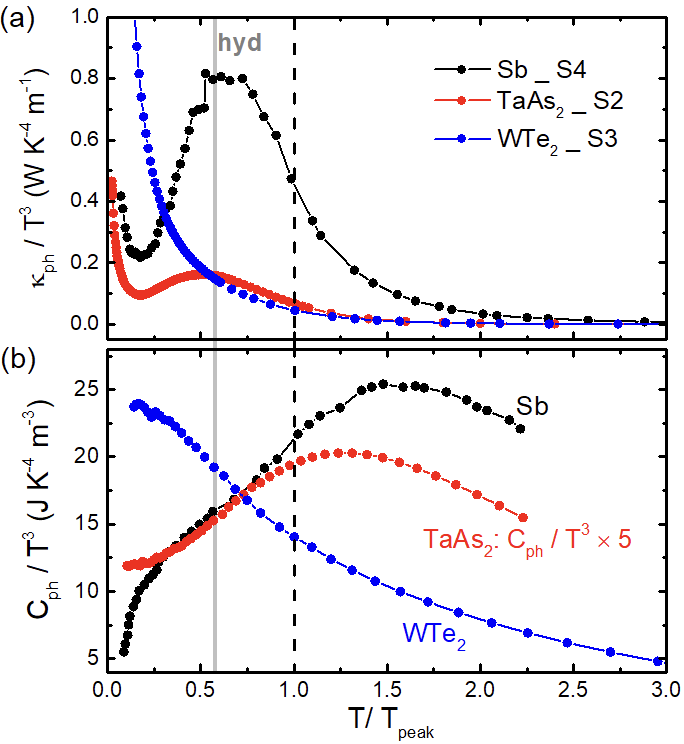}
\caption{\textbf{Phonon hydrodynamics and phonon anharmonicity.}
(b) $\kappa_{\mathrm{ph}}/T^3$ as a function of $T/T_{\mathrm{peak}}$ for Sb~\cite{Jaoui2022}, TaAs$_2$, and WTe$_2$~\cite{Xie2024}. $T_{\mathrm{peak}}$ corresponds to the $\kappa_{\mathrm{ph}}$ peak temperature, which is 9~K for Sb, 25~K for TaAs$_2$, and 16.5~K for WTe$_2$.
(b) $C_{\mathrm{ph}}/T^3$ as a function of $T/T_{\mathrm{peak}}$. Phonon hydrodynamics is related to phonon anharmonicity, as indicated by the upward deviation of $C_{\mathrm{ph}}$ from Debye theory.}
\label{figs6}
\end{figure*}

\end{document}